\documentstyle[12pt]{article}
\newcommand{\pr}{\paragraph{}}
\newcommand{\be}{\begin{equation}}
\newcommand{\ee}{\end{equation}}
\newcommand{\bea}{\begin{eqnarray}}
\newcommand{\nn}{\nonumber}
\newcommand{\eea}{\end{eqnarray}}
\newcommand{\nd}[1]{/\hspace{-0.6em} #1}
\newcommand{\nk}{\noindent}
\begin{document}
\begin{titlepage}
\vspace{.1in}
%
%

%

%
%
\begin{centering}
\vspace{.1in}
{\Large {\bf Valleys in Non-Critical String Foam Suppress
Quantum Coherence  }} \\
\vspace{.2in}
{\bf John Ellis$^a$}, {\bf N.E. Mavromatos$^{b}$} and {\bf D.V.
Nanopoulos}$^{a,c}$   \\
\vspace{.05in}
\vspace{.1in}
{\bf Abstract} \\
{\small
\paragraph{}
As an example of our non-critical string approach to
microscopic black hole dynamics,
we exhibit some string contributions to the $\nd{S}$ matrix
relating in- and out- state density matrices
that do not factorize as a product of $S$ and
$S^\dagger$ matrices. They are associated
with valley trajectories between
topological defects on the string world sheet,
that appear as quantum fluctuations
in the space-time foam. Through their
ultraviolet renormalization scale dependences
these valleys cause non-Hamiltonian
time evolution and
suppress off-diagonal entries
in the density matrix at large times. Our approach
is a realization of previous formulations
of non-equilibrium quantum statistical mechanics
with an arrow of time.}
\end{centering}
{\small
}
\par
\vspace{0.4in}
\vspace{0.1in}

\par
\vspace{0.5in}
%

%

\vspace{.2in}
--------------------------------------------- \\
$^a$ Theory Division, CERN, CH-1211, Geneva, Switzerland,  \\
$^b$ Laboratoire de Physique Th\`eorique
ENSLAPP (URA 14-36 du CNRS, associe\`e \`a l' E.N.S
de Lyon, et au LAPP (IN2P3-CNRS) d'Annecy-le-Vieux),
Chemin de Bellevue, BP 110, F-74941 Annecy-le-Vieux
Cedex, France, \\
on leave from P.P.A.R.C. Advanced Fellowship, Dept. of Physics
(Theoretical Physics), University of Oxford, 1 Keble Road,
Oxford OX1 3NP, U.K.  \\
$^c$ Center for Theoretical Physics, Dept. of Physics, \\
Texas A \& M University, College Station, TX 77843-4242, USA,
and \\
Astroparticle Physics Group,
Houston Advanced Research Center (HARC),
The Mitchell Campus, The Woodlands, TX 77381, USA\\

\end{titlepage}
\newpage
\section{Introduction}
\pr
Studies of quantum gravity
suggest \cite{bek,hawk2}
that a pure state description
cannot be maintained
in the context
of point-like field theories. It has been proposed \cite{hawk,ehns}
that quantum states be described by density matrices within
a framework that allows pure states to evolve
into mixed states with entropy increasing monotonically.
The transitions between asymptotic in- and out- states
would be governed by a superscattering matrix
$\nd{S}$ that does not factorize as a product of $S$ and $S^\dagger$
matrices as in point-like field theory \cite{hawk}.
The corresponding
evolution of
density matrices through intermediate times would include \cite{ehns}
a non-Hamiltonian term $\nd{\delta H}$ :
\be
\partial _t \rho = i[\rho, H] + \nd{\delta H}\rho
\label{one}
\ee
due to quantum fluctuations in the space-time
foam.
\pr
Although their failures
to solve the problem of quantum gravity
suggest the need
to modify quantum field theory and quantum mechanics,
point-like field theories
are inadequate to prove this, or
model $\nd{S}$ or $\nd{\delta H}$.
For this one needs a consistent quantum theory of gravity,
for which the only available
candidate is string theory.
No one questions the
validity of
quantum field theory and quantum mechanics on the world sheet.
Moreover, we have identified \cite{emn1}
an {\it infinite} {\it  set} of intrinsically
stringy symmetries sufficient to
preserve quantum coherence in the presence
of a space-time background with a horizon and/or
a singularity, such as a two-dimensional or (we conjecture)
spherically-symmetric
four-dimensional black hole. These $W$ symmetries were first identified
on the world sheet, couple string states at different string levels, and
are elevated to become local space-time gauge symmetries
with an {\it infinite set} of associated commuting
conserved charges available
to encode information\cite{emn1}. The $W$ charges
in target space are defined by {\it non-local} integrals
over space-time. For this reason, and because
laboratory experiments
observe only light particles from the lowest string level,
these cannot take $W$ symmetries and extended massive string
modes
into account.
Therefore we claim \cite{emnqm} that also in
string theory the quantum field theory
and quantum mechanics of the effective light-particle
theory must indeed be modified so as to accommodate transitions
from pure to mixed states, where the information
associated with massive modes is lost.
\pr
The form of the modifications can be derived \cite{emnqm}
by observing
that, although the full string theory is finite,
the truncated light-particle theory can be defined
only by introducing a renormalization scale cut-off, that
we identify with a negative-metric Liouville field
that can be interpreted as
the target time \cite{emnqm,emnuniv}. Thus we
arrive at a non-critical string description
of the effective light-particle theory, whose
time
(renormalization scale) -dependence
in a fixed space-time background
includes the Hamiltonian dynamics of conventional
quantum mechanics, which yields asymptotically the $S$
matrix of conventional quantum field theory, as
shown in ref. \cite{emnlong,emnnew}. However,
when fluctuations in the space-time background,
such as back-reaction and quantum fluctuations
in the space-time foam, are taken into account,
additional scale (time) -dependences appear that we claim
contribute non-Hamiltonian terms $\nd{\delta H}$
in (\ref{one}) and non-factorizable terms in
$\nd{S}$ \cite{emnqm}.
\pr
In a previous paper \cite{emnqm}, we used
the language
of world-sheet $\sigma$-models and considered
dynamics in the $(g^i, p_j)$ phase space,
where the $g^i$ denote $\sigma$-model couplings, that can
be regarded as coordinates in a space
endowed with the Zamolodchikov metric $G_{ij}$, the
$p^j$ are the corresponding conjugate momenta, and the
$\beta ^i$ are the corresponding renormalization
group $\beta$-functions. We found that this
dynamics was described by an equation
of the form (\ref{one}), with a Hamiltonian
$H(g^i, p_j)$ derived from the Zamolodchikov $C$-function \cite{zam}
that serves as an effective $\sigma$-model action by a Legendre
transformation, and derived
an expression
for $\nd{\delta H}$ :
\be
\nd{\delta H} =\beta^i G_{ij} \frac{\partial \rho}{\partial p^j}
\label{two}
\ee
Our modified Liouville equation (\ref{one},\ref{two})
is an explicit realization \cite{emnerice} of previous
approaches to non-equilibrium quantum statistical
mechanics. It exemplifies the general $\Lambda$-transformation
theory of ref. \cite{misra}, with an arrow
of time, and has a
Lie-admissible structure \cite{santilli},
as a result \cite{ktorides} of the symmetry
of the Zamolodchikov metric $G_{ij}$ \cite{emnerice}.
This also guarantees \cite{emnnew}
its consistency with canonical
quantization conditions \cite{hojman}.
\pr
The formalism leading to
(\ref{two}) is analogous to that of the
Drude model of quantum friction that was used
in refs. \cite{vernon,cald} to discuss decoherence
in open quantum-mechanical systems due to couplings
to `environmental oscillators'. In our approach,
the corresponding couplings are required by $W$-symmetry,
the unobservable extended massive string modes take the place of
the `environmental oscillators', and the resulting
loss of quantum coherence is regarded as an {\it inevitable }
consequence of fluctuations in the space-time background.
\pr
In this paper, we exhibit explicit
contributions to $\nd{\delta H}$  (\ref{two}) associated
with two distinct classes of space-time fluctuations : the quantum
creation and annihilation of a microscopic black hole, and the
back-reaction of light matter particles on the space-time
foam. We work in the context of the $SL(2,R)/U(1)$
coset Wess-Zumino model on the world-sheet, that describes
a two-dimensional (spherically-symmetric four-dimensional)
black hole\cite{witt}. We have argued previously that the quantum
creation and annihilation of microscopic
black holes is modelled by the dipole phase of a
monopole-antimonopole gas \cite{ovrut,emndua}
on the world-sheet, which therefore provides
suitable framework for analyzing the corresponding
contribution to $\nd{\delta H}$. As we discuss in section 2,
the back-reaction of light particles on the black-hole metric
is described by instantons \cite{yung}
in the $SL(2,R)/U(1)$ model
that capture the effects of couplings to the unobservable
extended massive string modes \cite{emnlong}.
\pr
In section 3, we evaluate monopole
and instanton contributions to $\nd{S}$ and $\nd{\delta H}$
by considering absorptive parts \cite{mueller}
of world-sheet
correlation functions that are dominated \cite{khoze}
by valley trajectories \cite{yung1}
in monopole-antimonopole
and instanton-anti-instanton configurations
respectively. We exhibit
explicit valley trajectories, and show
that they make contributions to the scale (time) -dependences
of transitions between density matrices that are linear
in small anomalous dimensions, contributing
terms
to $\nd{\delta H}$
and hence $\nd{S}$. As we discuss in section 4,
these have the effect
of suppressing off-diagonal
terms in the target configuration space
representation of the
effective light-particle density matrix
at large times.
\pr
\section{String Black Hole Monopoles and Instantons}
\pr
The action of $SL(2,R)/U(1)$ coset Wess-Zumino
model \cite{witt}
describing a Euclidean black hole can be written
in the form
\be
S=\frac{k}{4\pi} \int d^2z \frac{1}{1+|w|^2}\partial _\mu {\bar w}
\partial ^\mu w + \dots
\label{three}
\ee
where the conventional radial
and angular coordinates $(r,\theta)$ are given
by $w=sinh r e^{-i\theta}$ and the target
space $(r,\theta)$ line element is
\be
ds^2=\frac{dwd{\overline w}}{1 + w{\overline w}}=dr^2+tanh^2rd\theta^2
\label{four}
\ee
The Euclidean black hole can be written
as a vortex-antivortex pair \cite{emndua}, which is a solution of the
following Green function equations on a spherical world sheet:
\be
\partial _z\partial _{\bar z} X_v =i\pi \frac{q_v}{2}
[\delta (z-z_1)-\delta(z-z_2)]
\label{five}
\ee
The world-sheet can also accommodate
monopole-antimonopole pairs \cite{emndua}, which are solutions of:
\be
   \partial _z  \partial _{\bar z} X_m =-\frac{q_m \pi} {2}
[\delta (z-z_1) -\delta (z-z_2)]
\label{six}
\ee
These are related to Minkowski black holes
with masses $\propto q_m$.
We believe that some essential features
of four-dimensional space-time foam
can be captured by studying
an  analogue two-dimensional
model, in which the above
vortex and monopole configurations
are both regarded as sine-Gordon deformations
of the effective action for the field $X\equiv \beta^{-\frac{1}{2}}
{\tilde X}$, where $\beta^{-1}$ is an effective
`pseudo-temperature': $\beta =\frac{3}{\pi (c-25)} $ in Liouville
theory. The partition function \cite{ovrut}
\bea
Z&=&\int D{\tilde X} exp(-\beta S_{eff}({\tilde X}) )  \nn \\
\nonumber
\beta S_{eff}&=&  \int d^2 z [ 2\partial {\tilde X}
{\overline \partial } {\tilde X} +  \frac{1}{4\pi }
[ \gamma _v\omega ^{\frac{\alpha}{2}-2}
(2 \sqrt{|g(z)|})^{1-\frac{\alpha}{4}}: cos (\sqrt{2\pi \alpha }
[{\tilde X}(z) + {\tilde X}({\bar z})]):   \\
& +&   (\gamma _v, \alpha,
{\tilde X}(z) + {\tilde X}({\bar z}) )
\rightarrow (
\gamma _m, \alpha ', {\tilde X}(z) - {\tilde X}({\bar z}))]]
\label{seven}
\eea
requires for its specification an angular ultraviolet
cut-off $\omega$ on the world-sheet with metric $g (z,{\bar z})$.
Here $\gamma_{v,m}$ are the fugacities for vortices and spikes
respectively, and $\frac{\alpha}{4} $ is the conformal
dimension $\Delta$. This deformed sine-Gordon theory
has a low-temperature
phase modelling four-dimensional space-time foam,
in which monopole-antimonopole pairs are bound
in dipoles as irrelevant deformations
with the conformal dimension
\be
      \Delta _m =\frac{\alpha _m}{4}=
      \frac{\pi\beta}{2}q_m^2 > 1
\label{eight}
\ee
A monopole-antimonopole pair corresponds to the creation
and anihilation of a microscopic black hole
in the space-time foam.
\pr
As shown in ref. \cite{yung},
the $SL(2,R)/U(1)$  Wess-Zumino coset model
describing a Euclidean black hole also has instantons
given by the holomorphic function
\be
 w(z)=\frac{\rho}{z-z_0}
\label{nine}
\ee
with topological charge
\be
 Q=\frac{1}{\pi}\int d^2z \frac{1}{1+|w|^2}[{\overline \partial}
{\overline w}\partial      w  - h.c. ]
 =-2 ln(a) + ~const
\label{ten}
\ee
where $a$ is an ultraviolet cut-off discussed later.
The instanton action on the world-sheet  also depends
logarithmically on the ultraviolet cut-off. As in the case of
the more familiar vortex configuration in the Kosterlitz-Thouless
model, this logarithmic divergence does not prevent
the instanton from having important dynamical effects.
The instanton-anti-instanton vertices take the form \cite{yung}
\be
V_{I{\overline I}}\propto -\frac{d}{2\pi}
\int d^2z \frac{d^2\rho}{|\rho|^4}
e^{-S_0} (e^{(\frac{k[\rho\partial {\overline w} + h.c. + \dots ]}
{f(|w|)}}+ e^{(\frac{k[\rho \partial w + h.c. + \dots]}{f(|w|)}})
\label{eleven}
\ee
introducing a new term into the effective
action. Making a derivative expansion
of the instanton vertex and taking the large-$k$
limit, i.e. restricting our attention
to instanton sizes $\rho \simeq a$, this new term
has the same form as the kinetic term in (\ref{three}),
and hence corresponds to a renormalization of
the effective level parameter in the
large $k$ limit:
\be
 k \rightarrow k - 2\pi k^2 d'
\qquad : \qquad
 d' \equiv d\int \frac{d|\rho|}{|\rho|^3}
\frac{a^{2}}{[(\rho/a)^2 + 1]^{\frac{k}{2}}}
\label{twelve}
\ee
If other perturbations are ignored,
the instantons are irrelevant deformations
and conformal invariance is maintained.
However, in the presence of ``tachyon'' deformations,
$T_0 \int d^2z {\cal F}_{-\frac{1}{2}, 0,0}^{c,c}$
in the $SL(2,R)$ notation of ref. \cite{chaudh},
there are extra logarithmic infinities
in the shift (\ref{twelve}), that are visible in the dilute
gas and weak-``tachyon''-field approximations.
In this case, there
is a contribution to the effective action of the form
\be
T_0\int d^2z d^2z'<{\cal F}_{-\frac{1}{2}, 0, 0}^{c,c}
(z,{\bar z}) V_{I{\overline I}} (z',{\bar z}')>
\label{tachdeform}
\ee
Using the
explicit form of the ``tachyon''
vertex ${\cal F}$
\bea
{\cal F}_{-\frac{1}{2},0,0}^{c,c}=
\frac{1}{\sqrt{1 + |w|^2}}\frac{1}
{\Gamma (\frac{1}{2})^2}
\sum_{n=0}^{\infty} \{ 2\psi (n+1)
-2\psi (n+\frac{1}{2}) + \nn \\
+ ln(1 + |w|^2) \} (\sqrt{1 + |w|^2})^{-n}
\label{expression}
\eea
given by $SL(2,R)$ symmetry
\cite{chaudh}, it is straightforward
to isolate a logarithmically-infinite contribution
to the kinetic term in (\ref{three}), associated
with infrared infinities on the world-sheet
expressible in terms of the world-sheet area $\Omega /a^2$,
the latter being
measured in units of the ultraviolet cut-off $a$
\cite{emnlong,emnnew},
\bea
\propto  T_0 \int d^2z' \int
\frac{d\rho}{\rho} (\frac{a^2}{a^2 + \rho^2})^{\frac{k}{2}}
\int d^2 z \frac{1}{|z-z'|^2}
\frac{1}{1 + |w|^2}
\partial _{z'} w(z')
\partial _{\bar z'} {\overline w}(z') + \dots \nn \\
\propto  T_0 ln \frac{\Omega}{a^2} \int d^2z'
\frac{1}{1 + |w|^2}
\partial _{z'} w(z')
\partial _{\bar z'} {\overline w}(z')
\label{analyticexp}
\eea
Such covariant-scale-dependent contributions
can be attributed to Liouville field dynamics, through
the ``fixed-area constraint'' in the Liouville path
integral \cite{DDK,kutasov}. The
zero-mode part
can be absorbed in a
scale-dependent shift of $k$\cite{emnlong},
which for large $k >>1 $ may be assumed to exponentiate:
\be
k_R\propto (\frac{\Omega}{a^2})^{(const). \beta ^I T_0 }
\label{thirteen}
\ee
where $\beta ^I$ is the instanton $\beta$-function \cite{yung}.
In ref. \cite{emnlong} we gave general arguments
and verified to lowest order that instantons represent
higher mode effects, enabling us to identify
$\beta^I =-\beta ^T $, where
$\beta^T$ is
the renormalization-group $\beta$-function of a
matter deformation of the black hole\footnote{Notice that this
implies that the matter $\beta$-function has to be computed
in a non-perturbative way, which is consistent with the
exact conformal field theory analysis of ref. \cite{chaudh}.}.
Notice that in (\ref{thirteen}) both the
infrared and the ultraviolet cut-off scales enter.
In the following we shall not distinguish between
infrared and ultraviolet cut-offs. The physical
scale of the system, which varies along a
renormalization group trajectory, is the dimensionless
ratio of the two, which is identified with the Liouville
field.
\pr
The separation of the full string theory
into an effective light-particle
theory and higher modes described here by
instantons entails a non-critical
description of the former. This we achieve
using the Liouville string formalism,
with the Liouville field identified
not only as a renormalization scale,
but also as time, as we discuss in more detail
later.
\pr
The change in $k$ and the associated change in the
central charge $c=\frac{3k}{k-2}-1$
and the black-hole mass
$M_{bh} \propto (k-2)^{-\frac{1}{2}}$
do not conflict with any general theorems.
An analogous instanton renormalization
of $\theta$ has been demonstrated \cite{pruisk}
in related $\sigma$-models
that describe the Integer Quantum Hall Effect (IQHE), discussed further
in section 4. Instanton renormalization of $k$ can also be seen
in the Minkowski black hole model of ref. \cite{witt}, defined
on a non-compact manifold
$SL(2,R)/O(1,1)$\footnote{The $\sigma$-model action
of such a theory contains\cite{yung},
in addition to the action (\ref{three}),
a total-derivative $\theta$-term
which
can be thought of as a deformation of the
black hole by an ``antisymmetric tensor''
background, which in two dimensions
is a discrete mode as a result of the abelian gauge symmetry.
Its Euclideanized version
has also instanton solutions of the form (\ref{nine}),
but with {\it finite} action,
which
induce
``Liouville''-time-dependent shifts
to $k$, prior to matter couplings.}.
In our case, as we have seen, the instantons
reflect
a shift of the central charge between the matter
and background sectors of a combined matter $+$ black hole theory,
in which the total
central charge is unchanged.
They correspond to
a combination of world-sheet deformation operators in
Wess Zumino model\cite{witt}:
the exactly marginal
operator $L_0^2 {\overline L}_0^2 $ and the irrelevant part
of the exactly marginal deformation $L_0^1 {\overline L}_0^1$,
which involves an infinite sum
of higher-level string operators \cite{chaudh}.
It is well known that the light matter (`tachyon')
perturbation is not by itself exactly marginal, but this property
can be enforced by including these higher-level
string state operators
\cite{chaudh}.
The previously-mentioned
fact that, at large
$k$, these operators rescale the black hole metric, can also be seen
from
their contributions to the action of the deformed
Wess-Zumino
$\sigma$-model after the gauge field integration \cite{chaudh},
\bea
S_\sigma \ni \int d^2z \{\partial r {\overline
\partial} r (1-2g csch^2 r -2g sech^2 r) + \nn \\
\partial \theta {\overline \partial}\theta
(sinh^2 r + 2g - \frac{(sinh^2 r + 2g)^2}{cosh^2 r + 2g})\}
\label{chlyk}
\eea
where $g$ is the coupling of the $L_0^2 {\overline L}_0^2 $
deformation.
Changing variables $cosh^2r + 2g \rightarrow cosh^2r $ in (\ref{chlyk})
one finds that to $O(g)$ the target space metric is rescaled by an
overall constant,
and thus such perturbations have
the same effect as the instanton.
Thus the
instanton represents the effects of higher string
modes that are related to each other and to massless excitations
by a $W$ symmetry.
Matrix elements of the full exactly
marginal light matter $+$ instanton operator have
no dependence on the ultraviolet cut-off $a$, but the separate
matter and instanton parts do depend on $a$, as we have
seen above.

\section{Valleys in String Foam}
\pr
We now consider the contributions of monopoles
and instantons to $\nd{S}$ matrix elements
giving transitions between a generic initial-state
density matrix $\rho^{A(in)}_B$      and
final-state density matrix $\rho^{C(out)}_D$. This is
described by an absorptive part of a world-sheet
correlation function
\bea
&~&\sum _{X_{out}}~_{in}<A|D,X>_{out}~_{out}<X,C|B>_{in} =  \nn \\
&=&\sum _{X_{out}}~_{in}<0|T(\phi(z_A)\phi(z_D))|X>_{out}
{}~_{out}<X|{\overline T}(\phi(z_C)\phi(z_B))|0>_{in} = \nn \\
&=&~_{in}<0|T(\phi(z_A)\phi(z_D))
{\overline T}(\phi(z_C)\phi(z_B)|0>_{in}
\label{fourteen}
\eea
Here we have used the optical theorem \cite{mueller}
on the world sheet,
which is valid because conventional quantum field theory,
and indeed quantum mechanics, remain valid on the world sheet,
to replace the sum over unseen states
$X$ by unity. Next, we use dilute-gas approximations
to estimate the leading monopole-antimonopole
and instanton-anti-instanton contributions to the absorptive part
(\ref{fourteen}). We expect these to be dominated \cite{khoze}
by valley configurations in the Euclidean functional
integral, so that in a semi-classical approximation
\be
\nd{S} \propto Abs\int D\phi _c exp(-S_v(\phi _c))F_{kin}
\label{fifteen}
\ee
where the integral is over the collective coordinates $\phi _c$
of the valley, whose action is $S_v(\phi _c)$. The function
$F_{kin}$ depends on kinematic factors, taking generically
the form
\be
F_{kin}=exp(E\Delta R)
\label{sixteen}
\ee
in the case of a four-point function for large $E\Delta R$,
where $E$ is the centre-of-mass energy and $\Delta R$ is the
valley separation parameter. This enables us to make a saddle-point
approximation to the integral (\ref{fifteen}), which we then
continue back to Minkowski space.
\pr
Valley trajectories $\psi _v$ have a homotopic parameter
$\mu$ and obey an equation of the form
\be
\frac{\partial S_0}{\partial \psi}|_{\psi =\psi_v}= W_\psi (\mu)
\frac{\partial \psi _v}{\partial \mu}
\label{seventeen}
\ee
where $W_{\psi} (\mu) $ is a weight function that
is positive definite
and decays rapidly at large distances \cite{yung1,khoze}. We
adapt techniques used in the $O(3)$ $\sigma$-model \cite{dorinst,BW}
to find the
monopole-antimonopole and instanton-anti-instanton valleys in a
reduced version of the $SL(2,R)/U(1)$ model. The separations
of the topological defects and anti-defects are well-defined in
the presence of conformal symmetry breaking, which is provided
in our case by the dilaton field \cite{emnlong}. Valleys can be
found by using the analogy \cite{dorinst}
between $\mu$ and a `time' variable
for defect-anti-defect scattering. We do not discuss here
the details of their construction, but record the results.
\pr
The monopole-antimonopole valley function,
expressed in terms of the original world-sheet
variables, reads
\be
 w(z,{\bar z})=\frac{(v-1/v){\bar z}}{1 + |z|^2}
\label{cocentric}
\ee
where $v$ denotes the separation in the $\sigma$-model
framework. Eq. (\ref{cocentric}) represents a
concentric valley, which can then be mapped into an
ordinary valley by applying appropriate conformal
transformations. The function (\ref{cocentric})
interpolates between a far-separated
monopole-antimonopole pair ($v \rightarrow \infty$)
and the trivial vacuum ($v=1$).
For large but finite
separations the corresponding
valley action leads to the action of a monopole-antimonopole
pair interacting via dipole interactions.
The action of the monopole-antimonopole valley
depends on the angular ultraviolet cut-off $w$ introduced
in section 2:
\be
S_m=8\pi q^2 ln (2)\sqrt{2}e^\gamma
+ 2\pi q^2 ln\frac{2R}{\omega  } +
2\pi q^2 ln[\frac{|z_1-z_2|}{(4R^2 + |z_1|^2     )^{\frac{1}{2}}}
\frac{4}{(4R^2 + |z_2 |^2     )^{\frac{1}{2}}}]
\label{eighteen}
\ee
for a monopole and antimonopole pair of equal
and opposite charges $q$, which we treat as a
collective coordinate over which we must integrate,
where $\gamma$ is Euler's constant, the second term
in (\ref{eighteen}) is a logarithmically-divergent
self-energy term on a spherical world sheet of
radius $R$, and the last term in
(\ref{eighteen}) is a dipole interaction energy.
For finite separations $0 < |z_1 -z_2 | < \infty $
and very small
world-sheets $R=O(a \rightarrow 0)$, the action (\ref{eighteen})
yields
\be
S_m=2\pi q^2 ln  \frac{a}{\omega  } + finite~parts
\label{nineteen}
\ee
where the ultraviolet cut-off dependence is apparent.
\pr
To construct the instanton valley, we notice that
in the reduced model used for the construction
of the monopole valley (\ref{cocentric})
the solution for an instanton-anti-instanton pair
is derived from the corresponding monopole case
via a conformal transformation in the $(\mu, ln|z|)$-plane.
In ref. \cite{emnlong} we give arguments why this construction
is true for finite separations as well, thereby leading
to an expression of the instanton valley as an (approximate)
conformal transform of the monopole valley (\ref{cocentric}).
The action of the instanton-anti-instanton
valley in the large-separation limit of the
dilute-gas approximation is
\be
S_{I{\overline I}}=kln(1 +|\rho |^2/a^2) +
O(\frac{\rho {\overline \rho}}{(\Delta R)^2)})
\label{twenty}
\ee
where $\Delta R$ is the separation of an instanton of
size $\rho$ and an anti-instanton of size ${\overline \rho}$,
and we find a dependence on the ultraviolet
cut-off $a$.
The actions (\ref{nineteen},\ref{twenty}) substituted
into the general expression (\ref{fifteen}) make non-trivial
contributions to the $\nd{S}$ matrix that do not
factorize as a product of $S$ and $S^\dagger$ matrix
elements.

\section{Suppression of Coherence}
\pr
To understand
how the above
non-perturbative contributions suppress
quantum coherence
at large times, we now review
our interpretation \cite{emnqm,emnlong,emnuniv}
of target time
as a renormalization group scale parameter in the
effective light-particle theory
with an ultraviolet cut-off (cf, $\omega$ in the
monopole case, $a$ in the instanton case).
We use the concept of the local (on the world sheet) renormalization
group equation \cite{shore,osborn}, according
to which perturbative $\beta$-functions
constitute a
gradient flow of
the string effective action in target space \cite{santos},
and
identify the
local cut-off as the Liouville mode, whose kinetic term
has a temporal signature for supercritical strings
($c > 26)$ \cite{aben}. The
inclusion of topological fluctuations such as monopole-antimonopole
or instanton-anti-instanton pairs in a critical string
theory makes it supercritical
{\it locally}, necessitating
the introduction of such a time-like Liouville field,
which we interpret as target time \cite{emnqm,emnlong,emnuniv}.
\pr
We
regard the two-dimensional black hole model of ref. \cite{witt}
as a toy laboratory that gives us insight into the nature
of time in string theory and contributes to the physical effects
that will be of interest to us here.
The action of the model is
\be
   S_0=\frac{k}{2\pi} \int d^2z [\partial r {\overline \partial } r
- tanh^2 r \partial t {\overline \partial } t] + \frac{1}{8\pi}
\int d^2 z R^{(2)} \Phi (r)
\label{action}
\ee
where $r$ is a space-like coordinate and $t$ is time-like,
$R^{(2)}$ is the scalar curvature, and $\Phi$ is the dilaton field. The
customary interpretation of (\ref{action}) is as a string model with
$c$ = 1 matter, represented by the $t$ field, interacting with a
Liouville mode, represented by the $r$ field, which has $c  < 1$ and
is correspondingly space-like \cite{aben,mm,aben3}.
As an illustration of the
approach outlined above, however, we re-interpret
(\ref{action}) as a fixed point of the renormalization group
flow in the local scale variable $t$. In our interpretation, the
``matter'' sector is defined by the spatial coordinate $r$, and has
central charge $c_m$ = 25 when $k  = 9/4$
\cite{witt}. Thus the model
(\ref{action}) describes a critical string in a dilaton/graviton
background. The fact that this is static, i.e. independent of $t$,
reflects the fact that one is at a fixed point of the renormalization
group flow \cite{emnlong,emnuniv}.
\pr
We now outline how one can use the machinery
of the renormalization group in curved space,with $t$ introduced as
a local renormalization
scale on the world sheet, to derive the model (\ref{action}). A detailed
technical description is given in \cite{emnlong,emnuniv,emnnew}.
There are two contributions
to the kinetic term for $t$ in our approach, one associated with
the Jacobian of the path integration over the world-sheet metrics, and
the other with fluctuations in the background metric.
\pr
To exhibit the former, we first choose the conformal gauge
$\gamma _{\alpha \beta}=e^{\rho}
{\hat \gamma}_{\alpha\beta}$ \cite{DDK,mm},
where $\rho$ represents the Liouville mode. One must identify $\rho$
with 2$\alpha'  \phi$, where $\phi (z, {\bar z}) \equiv ln \mu
(z, {\bar z})$ is a local scale on the world-sheet,
in order to
reproduce the critical string results of ref. \cite{witt}.
This makes the local scale
$\phi$ a dynamical $\sigma$-model field. Ref.\cite{mm}
contains an
explicit computation of the Jacobian using heat-kernel regularization,
which yields
\be
-\frac{1}{48\pi}[\frac{1}{2}
\partial_\alpha \rho \partial^\alpha \rho +
R^{(2)}\rho + \frac{\mu}{\epsilon} e^\rho +
S'_G ]
\label{liouvillekin}
\ee
where the counterterms $S'_G$ are needed to remove the non-logarithmic
divergences associated with the induced world-sheet
cosmological constant term $\frac{\mu}{\epsilon}e^\rho$,
and depend on the background fields.
\pr
To find the contributions to the effective
action from the background fields
we recall that the renormalization of composite operators in
$\sigma$-models formulated on curved world sheets is achieved by
allowing an arbitrary dependence of the couplings $g^i$ on the
world-sheet variables $z,{\bar z}$ \cite{shore,osborn}.
This induces counterterms of ``tachyonic'' form,
which, in dimensional regularization with
$d  = 2 -  \epsilon$,
include the following simple pole at one loop, for a
$\sigma$-model propagating in a graviton background
\cite{osborn}:
\be
Y^{(1)}=\frac{\lambda}{16\pi \epsilon}\partial _\alpha G_{MN}
\partial^\alpha G^{MN}
\label{polemetric}
\ee
where $\lambda\equiv 4\pi \alpha '$ is a loop-counting parameter.
In ref. \cite{osborn}
$G_{MN}$ was allowed to depend arbitrarily on the world-sheet
variables, and all world-sheet derivatives of the couplings were
set to zero at the end of the calculation. In our Liouville mode
interpretation, we assume that such dependence occurs only through
the local scale
$\mu (z,{\overline z})$,
so that
\be
\partial _\alpha g^i ={\hat \beta^i}\partial _\alpha \phi (z,{\bar z})
\label{part}
\ee
where ${\hat \beta }^i=\epsilon g^i + \beta ^i(g)$
and $\phi=ln \mu (z,{\bar z})$.
Taking
the $\epsilon \rightarrow 0$ limit,
separating the finite and $O(\frac{1}{\epsilon})$ terms,
and taking into account that
the one-loop graviton $\beta$-function
is
\be
\beta _{MN}^G=\frac{\lambda}{2\pi}R_{MN}
\label{gravbeta}
\ee
and that
in the case of the Minkowski black hole model of ref. \cite{witt},
the
Lorentzian target-space curvature is
\be
R=\frac{4}{cosh^2r}=4-4tanh^2r,
\label{curba}
\ee
we recover, upon
combining the world-sheet metric Jacobian term in (\ref{liouvillekin})
with the gravitational
background fluctuation terms,
the critical string $\sigma$-model action
(\ref{action})
for the Minkowski black-hole.
Dilaton counterterms are incorporated
in a similar way, yielding the dilaton
background of \cite{witt}. In addition,
as standard in stringy
$\sigma$-models, one also obtains the necessary
counterterms that guarantee target-space diffeomorphism
invariance of the Weyl-anomaly cefficients \cite{shore}.
Details are given in ref. \cite{emnnew}.
\pr
It should be noticed that
the renormalization group yields automatically the Minkowski
signature, due to the $c_m =25$ value of the matter central charge
\cite{aben,aben3}.
However, the relevant
conformal field theory is more developed
in the Euclidean case \cite{chaudh}.
As we
remarked in ref. \cite{emnnew,emnuniv,emnlong},
one can switch over
to the Euclidean black hole model, and still maintain the
identification of the compact time with some appropriate function
of the Liouville scale $\phi$  that takes into account the
compactness of $t$ in that case.
In ref. \cite{chaudh} it was argued that the exactly marginal
deformation that turned on a static tachyon background for the
black hole of ref. \cite{witt} necessarily involved the
higher-level topological string
modes, which are non-propagating
extended
states analogous to the de-phasons of the Hall model \cite{libby}.
This is a consequence
of the operator product expansion of the tachyon zero-mode operator
${\cal F} _{-\frac{1}{2},0} ^c    $ \cite{chaudh}:
\be
    {\cal F} _{-\frac{1}{2},0}^c    \circ
    {\cal F} _{-\frac{1}{2},0}^c    = {\cal F}_{-\frac{1}{2},0}^c
+ W_{-1,0}^{hw} + W_{-1,0}^{lw} + \dots
\label{ope}
\ee
where we only exhibit
the
appropriate holomorphic part for reasons of economy of
space.
The
$W$ operators and the $\dots$ denote level-one and higher string
states. The latter cannot be detected in local scattering
experiments, due to their extended character.
{}From a formal field-theoretic point of view such states cannot
exist as asymptotic states to define scattering, and also cannot
be integrated out in a local path-integral. They can only
exist as marginal deformations in a string theory.
An `experimentalist' therefore sees necessarily a
truncated matter theory, where the only deformation  is the
tachyon ${\cal F}_{-\frac{1}{2},0}^c    $, which is a (1,1) operator
in the black hole $\sigma$-model (\ref{action}), but is not
exactly marginal. This implies additional Liouville dressing,
due to the non-vanishing operator-product-expansion coefficients.
The latter contribute to the quadratic, and higher order, parts of
the Wilson Renormalization Group $\beta$-functions of the
pertinent $\sigma$-model which are
non-zero.
Hence, this truncated theory
is non-critical, and the additional
Liouville dressing
in the sense is essential, implies
time-dependence of the matter background.
Due to the
fact that the
appropriate
 exactly-marginal deformation associated with the tachyon in
these models
 involves all higher-level
string states, one can
conclude that in this picture the ensuing
non-equilibrium time-dependent backgrounds
are a consequence
of information carried off
by the unobserved topological string modes.
The world-sheet instantons
studied in section 2, which in the presence of tachyons
lead to violations of conformal invariance,
is a way to represent {\it collectively}, in a semi-classical
approximation, the effects of
these topological string modes on the low-energy
field theory.
The r\^ole of the space-time singularity\footnote{We would like to
stress that the notion of `singularity' is clearly a low-energy
effective-theory concept. The existence of infinite-dimensional
stringy symmetries associated with higher-level string states
($W_\infty$-symmetries \cite{emn1}) `smooth out' the singularity,
and render the full string theory finite.}
 was crucial for this argument.
Indeed, in flat target-space matrix models \cite{matrix} the
tachyon zero-mode operator ${\cal F}$ is exactly marginal. As we
argue in ref. \cite{emn1,emnuniv},   these flat models can be regarded as
an asymptotic ultraviolet
limit in time
of the Wess-Zumino black hole. Hence,
any time-dependence of the matter disappears in the vacuum,
which is thus viewed as an `equilibrium' situation for the string.
\pr
There is
a very important aspect of Liouville theory that
makes it distinct from any other approach with
a local renormalization group cut-off.
When evaluating correlation functions
among Liouville-dressed vertex operators, the
result is proportional \cite{li}
to $\Gamma (-s)$ factors,
where $s$ is the sum of the respective
Liouville energies, including the terms
of the background
charge at infinity that represent the Liouville mode itself.
Among the physically-interesting values of $s$ are those
corresponding to positive integers, which for
instance is the case of light-matter scattering off
a string black hole,
that involves an excitation of the latter
to a discrete massive state. Such values of $\Gamma (-|n|)$
call for regularization via analytic continuation \cite{kogan2}
along a saddle-point contour of the form illustrated
in the figure. The result is
\be
    \Gamma (-|n|) =\int _C dA A^{-|n|-1} e^{-A}
\label{contour}
\ee
implying the existence of imaginary parts, which are then
interpreted as indicating the renormalization
group instability.
By inspection of the figure and applying the logic
of ref. \cite{coleman}, we observe that
the effective flow of target time
is from an infrared fixed point with large
apparent world-sheet area in target space, to an ultraviolet
fixed point
with small world-sheet area. This flow is analogous to the
decay
of a metastable vacuum in conventional
field theory.
It is characterized by {\it energy conservation} \cite{emncpt}
and a
{\it monotonic increase in entropy} \cite{emnqm}.
\pr
In the dilute-gas approximation introduced
in section 3, the topologically trivial zero
monopole-antimonopole, zero instanton-anti-instanton
sector in the unitary sum in (\ref{fourteen}) provides
the usual $S$-matrix description
of scattering in a fixed background, with no
back reaction of the light matter on the metric.
This result is well-known in the
conformal field theory approach to critical
string theory, and is discussed explicitly
in the present context in section 6 of ref. \cite{emnlong}.
This $S$-matrix contribution corresponds to the usual
Hamiltonian description of quantum mechanics, via the
representation $S=1 +i T$ : $T=\int _{-\infty}^{\infty}
dt H(t)$. Any topologically non-trivial contribution
to the unitarity sum in (\ref{fourteen}) goes beyond
the usual treatment of conformal field theory in critical strings,
and
makes a contribution
to the non-factorization of the $\nd{S}$-matrix : $\nd{S}=S S^\dagger
+ \dots$. Two such contributions that we have identified above
come from the monopole-antimonopole and instanton-anti-instanton
sectors discussed above, which we expect to be dominated by the
valley actions (\ref{nineteen}) and (\ref{twenty}) respectively.
\pr
The dependence of the monopole-antimonopole valley
action (\ref{nineteen}) on the ultraviolet
cutoff $\omega$,
which we identify with the target-space time $t=-ln\omega$,
and of the instanton-anti-instanton valley action (\ref{twenty})
on the local scale-dependent level parameter $k$ (\ref{twelve},
\ref{thirteen}) where $t=-ln a$, tell us that both
valleys contribute to the non-Hamiltonian term in
the modified quantum Liouville equation (\ref{one}).
{}From
the monopole-antimonopole valley (\ref{nineteen})
we find \cite{emnlong}

\be
\Delta \nd{S} \simeq e^{-2(\Delta _m -1) t + \dots }
\label{twentythree}
\ee
where $\Delta _m -1 $ is the anomalous dimension
(\ref{eight}) of an irrelevant dipole-like monopole-antimonopole
pair, and from the instanton-anti-instanton valley (\ref{twenty})
we find \cite{emnlong}
\be
\Delta \nd{S} \simeq e^{-2\gamma _0 t + \dots }
\label{twentyfour}
\ee
where $\gamma _0$ is the anomalous dimension
(\ref{thirteen}) of a matter deformation. Both
of these time-dependences apply in limits of far-separated
defect and anti-defect. Differentiating (\ref{twentythree},
\ref{twentyfour}) with respect to time, we find contributions
to $\nd{\delta H}$ that are proportional to the
anomalous dimensions $(\Delta _m -1),\gamma _0$.
As we discussed in ref. \cite{emnnew}, it is a
feature of our non-critical Liouville string approach
that correlation functions in general, and $\nd{S}$
in particular, receive area-  and hence time-dependences
proportional
to the anomalous dimension of any deformation.
\pr
The time-irreversibility
inherent in(\ref{twentythree},\ref{twentyfour})
reflects the fact that our modified
Liouville equation (\ref{one},\ref{two})
possesses an arrow of time \cite{emnerice}.
Moreover, it has a Lie-admissible
structure \cite{santilli},
as a consequence
\cite{ktorides}
of the symmetry
of the Zamolodchikov metric $G_{ij} $ \cite{emnerice},
and is consistent
\cite{emnnew} with canonical
quantization \cite{hojman}, again as a result
of the symmetry of $G_{ij}$ and the gradient
property of the renormalization group flow.
\pr
The resulting formalism for the
time evolution of the density matrix is analogous
to the Drude model of quantum
friction \cite{vernon,cald}, with the
massive string modes playing the r\^oles of
`environmental oscillators'. In the language of
world-sheet $\sigma$-model couplings $\{ g \}$,
the reduced density matrix of the observable states is given, relative
to that evaluated in conventional Schr\"odinger quantum mechanics,
by an expression of the form
\be
 \rho (g, g',  t  ) / \rho_S (g, g',  t  )
\simeq
e^{-\eta \int_0^{t} d\tau \int_{\tau ' \simeq \tau }
d\tau '
 \beta ^i G_{ij} \beta ^j } \simeq
e^{ - Dt ({\bf g}  - {\bf  g'} )^2 + \dots }
\label{twentyone}
\ee
where $\eta$ is a calculable proportionality coefficient,
and $G_{ij}$ is the Zamolodchikov metric \cite{zam}
in the space
of couplings. In string theory, the identification of the target-space
action with  the Zamolodchikov $c$-function $C(\{g \})$ \cite{zam}
enables the Drude exponent to be written in the form
\be
\beta^iG_{ij}\beta^j = \partial _t C(\{g\})
\label{twentytwo}
\ee
which also determines the rate of increase
of entropy
\be
{\dot S}=\beta^i G_{ij}\beta^j S
\label{twentytwob}
\ee
In the string analogue (\ref{twentytwo}) of the Drude model
(\ref{twentyone}) the r\^ole of the coordinates in (real) space
is played by the $\sigma$-model couplings $g^i$ that are
target-space background fields. Relevant for us
is the tachyon field $T(X)$, leading us to interpret
$(g-g')^2$ in (\ref{twentyone}) as \cite{emnlong}
\be
(g-g')^2=(T-T')^2 \simeq (\nabla T)^2 (X-X')^2
\label{new}
\ee
for small target separations $(X,X')$.
\pr
The effect of the time-dependences (\ref{twentythree},\ref{twentyfour},
\ref{twentyone},\ref{new}) is
to suppress off-diagonal elements
in the target configuration space
representation \cite{emohn}of the out-state
density matrix :
\be
\rho _{out} (x,x') ={\hat \rho }(x)\delta (x-x')
\label{twentyfive}
\ee
This behaviour can be understood intuitively
as being related to the apparent shrinking of the string
world sheet in target space, which destroys interferences
between strings localized at different points in target
configuration space, c.f.
the de-phasons in the Hall model \cite{libby}. This
behaviour is generic for
string contributions to the space-time foam, which
make the theory supercritical locally, inducing renormalization
group (target time) flow. The two specific
contributions (\ref{twentythree},\ref{twentyfour})
to this suppression (\ref{twentyfive}) of space-time
coherence that we have identified
in this paper correspond (\ref{twentythree})
to microscopic black
hole formation and (\ref{twentyfour}) to the back-reaction
of matter on a microscopic black hole, entailing in each case
information loss across an event horizon.
\pr

\nk {\Large{\bf  Acknowledgements}}
\pr
The work of N.E.M. is supported by a EC Research Fellowship,
Proposal Nr. ERB4001GT922259.
That of D.V.N. is partially supported by DOE grant
DE-FG05-91-GR-40633.
J.E.
would like to thank LAPP (Annecy-le-Vieux,
France) for its hospitality while part of this work
was being done. We would also like to thank Tim Hollowood
and A.V. Yung
for useful discussions.

\newpage
\pr
{\Large {\bf Figure Caption}}
\pr
\nk Contour of integration in the analytically-continued
(regularized) version of $\Gamma (-s)$ for $ s \in Z^+$.

\end{document}